# How to compute the atomic stress objectively?

Bin Liu[1,2,*], Xinming Qiu[1,*]

[1] *FML, Department of Engineering Mechanics, Tsinghua University, Beijing 100084, China*

[2] *State Key Laboratory of Structural Analysis for Industrial Equipment, Dalian University of Technology, Dalian 116023, P.R. China*

[*]Corresponding authors. Tel.: 86-10-62786194; fax: 86-10-62781824

E-mail address: liubin@tsinghua.edu.cn (B. Liu),

qxm@tsinghua.edu.cn (X. Qiu)

**Abstract**

Atomistic simulation has been a powerful study tool in mechanics research, but how to objectively compute the atomic stress equivalent to Cauchy stress is still controversial, especially on the velocity-related part in the virial stress definition. In this paper, by strictly following the classical definition of the Cauchy stress for continuum medium, the fundamental Lagrangian atomic stress is proposed and can be used to obtain the correct Cauchy stress under any circumstances. Furthermore, the Lagrangian virial stress is proposed, which is still in virial form but does not include velocities to avoid controversial velocity treatments. It is also found that the widely used classical virial stress is actually the Eulerian virial stress, which includes the velocities of atoms, and is valid only when the impulse-momentum theorem is applicable to estimate the internal forces. However this requirement for the Eulerian atomic stress can not always be met in practical cases, such as the material volume element in rotation and the examples presented in this paper, but the proposed Lagrangian atomic stress can avoid these velocity-related nonobjectivities.

## 1. Introduction

Continuum mechanics has been very successful in predicting material behaviors at macroscopic scale. Moreover, with the emergence of nanotechnology and nanoscience, many recent researches demonstrated that, in many cases, the concepts of continuum mechanics can still be applied to discrete atom systems at microscopic scale (e.g., Refs.[1-5]). On the other hand, atomistic simulation has been a powerful study tool in mechanics research. In the theoretical framework of continuum mechanics, Cauchy stress is one of the most important quantities. How to correctly extract stress information from atomistic simulations is a key to link atomistic and continuum studies. For discrete atom systems, the virial stress and its modified editions,[6-11] which consist of both kinetic and potential parts, have been widely used in atomistic simulations and believed to be equal to its continuum counterpart, the Cauchy stress. However, Zhou[12] argued that only the potential part corresponds to the Cauchy stress since the stress is only the measure of average internal forces among atoms and should be independent of the velocities of atoms. Zhou's work[12] has attracted many attentions and led to a lot of confusions, and many researchers have used this stress definition in their works (e.g., Refs[13-16]). But some theoretical and numerical works[17,18] still support the virial stress as the Cauchy stress, so this controversy has not been settled yet. In the following, we use two examples to demonstrate the uncertainness or incorrectness of the definition of the classical virial stress.

The first example is a dynamical problem of a solid bar as shown in Fig. 1. The bar is initially stretched, then the stretching forces are suddenly released and the bar begins vibrating. The molecular dynamics is used to simulate this problem, and we want to compute the stress at point $A$ at the moment $t=t_1$. It is found that the existing atomic stress definition can not provide the consistent stress value. If the classical virial definition

$$\sigma = \frac{1}{V}\left(-\sum_i m_i \mathbf{v}_i \otimes \mathbf{v}_i + \frac{1}{2}\sum_{i,j\neq i} \mathbf{r}_{ij} \otimes \mathbf{f}_{ij}\right)$$

is used, the stress will be different for the different reference frames (i.e., two observers with different velocities) since the velocity $\mathbf{v}_i$ will be different while the potential part $\frac{1}{2}\sum_{i,j\neq i} \mathbf{r}_{ij} \otimes \mathbf{f}_{ij}$ keeps the same. The other popular virial stress definition is

$$\sigma = \frac{1}{V}\left(-\sum_i m_i (\mathbf{v}_i - \bar{\mathbf{v}}) \otimes (\mathbf{v}_i - \bar{\mathbf{v}}) + \frac{1}{2}\sum_{i,j\neq i} \mathbf{r}_{ij} \otimes \mathbf{f}_{ij}\right),$$

where $\bar{\mathbf{v}}$ is the average velocity of the atoms in the volume element. However, for this example, different volume elements will lead to the different stresses. If the whole bar is used as the volume element, the average velocity $\bar{v} = \boldsymbol{0}$; if the dashed box shown in the figure is used as the volume element, the average velocity $\bar{v} \neq \boldsymbol{0}$. This inconsistence makes us realize that the velocity is not an objective quantity, which depends on the choice of the reference frame or the volume element.

Moreover, from the second example shown in Fig. 2, it can be found that the velocity also depends on the external forces. Figure 2 shows a box consisting of some charged atoms or ions, and the distances among them are very large such that there is no interatomic interaction. If an alternative electric field $E$ is applied to this system and the amplitude and frequency of $E$ can be adjusted, one can imagine these charged atoms vibrate locally and the magnitude of the velocities can be very large. Therefore, the kinetic part of virial stress has negative components and the potential part is vanishing such that the virial stress also has negative components. By contrast, since the temporal average external force on any material point is zero and the system is obviously in thermodynamic equilibrium, the temporal average Cauchy stress should be zero according to the balance of forces. However, it is easy to note that the temporal average classical virial stress is not vanishing at all. This obvious inconsistence between Cauchy stress and the classical virial stress further indicates that the classical virial stress definition is not widely applicable due to the involvement of the velocity which relies on too many factors.

In this paper, we will start from the fundamental definition of the Cauchy stress for continuum medium to derive its counterpart in atomic systems, and then discuss what definition of atomic stress is objective for discrete atom systems. We noted that there have already been many elegant and mathematical ways to derive the virial stress and other definitions of atomic stress. However, we think that these different lengthy derivations and definitions, such as Zhou's work[12], might confuse the readers. Alternately, the strategy of this paper is using a direct and easy way to obtain various definitions on the atomic stress, and using many examples to demonstrate their limitations or incorrectness.

**2. Definition of Cauchy stress for continuum medium**

The following definition of Cauchy stress can be found in many textbooks (e.g., Ref.[19]). A continuum body is divided by a plane through point $P$ as shown in Fig. 3. The internal force $f$ between two parts are then exposed. The Cauchy stress tensor $\sigma$ of point $P$ is defined through

$$\boldsymbol{n} \cdot \boldsymbol{\sigma} = \frac{\boldsymbol{f}}{S}, \qquad (1)$$

where $\boldsymbol{n}$ is the unit normal vector of the dividing plane and $S$ is the cross-sectional area. As pointed out by Zhou[12], this definition is valid for all scenarios, fully dynamic and static.

It is important to note that this dividing plane is a Lagrangian cut, or material cut. Once the dividing is made, the material points are separated into two sets, and their identities will not change hereafter. Only the internal forces between the two sets are counted in $\boldsymbol{f}$. In addition, the Cauchy stress in continuum medium is actually a statistic average over a spatial and temporal dimension in which the noisy fluctuation at the atomic scale can be smeared out.

The Cauchy stress has been proved to satisfy the momentum balance law and can be successfully used in continuum mechanics. In the following, we will follow the same way to define the atomic stress for discrete systems. It should be pointed out that although there are many stress definitions in continuum mechanics, the goal of this paper is seeking for a proper definition of atomic stress which is equivalent to the Cauchy stress.

### 3. Fundamental Lagrangian atomic stress

Figure 4 shows an illustrative volume element in a discrete atom system, which should be macroscopically small for determining local information and microscopically large with a great number of atoms to conduct statistically meaningful average. Adopting the previous treatment in continuum mechanics, we divide this element into two sets (Set L and Set R) by a material cut, and mark the atoms on each side with different colors in Fig. 4, white for the left and black for the right. It should be emphasized that this cut is a Lagrangian (or material) dividing plane, namely each atom will not change its belongingness to Set L or Set R. Similarly, with the spatial and temporal average over the left set, the definition on the atomic stress $\sigma$ then is,

$$\boldsymbol{n} \cdot \boldsymbol{\sigma} = \left\langle \frac{1}{S} \sum_{i \in Lag\ Set\ L} \boldsymbol{f}_{iR}^{Lag} \right\rangle, \qquad (2)$$

where $\langle \cdot \rangle$ denotes the temporal average, $\boldsymbol{f}_{iR}^{Lag}$ is the force on atom $i$ exerted by the atoms in Lagrangian Set R, $S$ is the cross-sectional area of the volume element and $\boldsymbol{n}$ is the unit normal vector of the dividing plane. We name the stress $\boldsymbol{\sigma}$ in Eq. (2) the *fundamental Lagrangian atomic stress* hereafter. Obviously, it is equivalent to the Cauchy stress because both of their definitions follow the same approach. It is noted that this fundamental Lagrangian atomic stress in Eq. (2) does not include the velocities of atoms, and is independent of the velocity of the reference frame and therefore is objective. More discussions on objectivity will be given later in this paper.

It should be emphasized that the temporal average in Eq. (2) is necessary and can not be neglected. Actually, the ignorance of the temporal average in Eq. (2) essentially yields the definition on the atomic stress $\hat{\sigma}$ suggested by Zhou[12],

$$\boldsymbol{n} \cdot \hat{\boldsymbol{\sigma}} = \frac{1}{S} \sum_{i \in Set\ L} \boldsymbol{f}_{iR} . \qquad (3)$$

Equation (3) is only dependant on the quantity at the moment $t$, so the Lagrangian set here is the same as the Eulerian one and "Lag" is therefore omitted. We use a simple example to demonstrate its incorrectness. Figure 5 shows a snapshot of an ideal atom system consisting of four one-dimensional atomic chains with harmonic interatomic potential at thermal and mechanical equilibrium (no mechanical loading is assumed here). The atoms in the first and third chains are at their equilibrium positions at this moment, and those in the second and fourth ones deviate from their corresponding equilibrium positions and are not uniformly distributed — some of them move closer with compressed bonds (colored blue), some separate with tensile bonds (colored red). Since this system is in equilibrium and not subject to any mechanical loading, $\sigma_{xx}$ should be zero due to the balance of momentum. If only the spatial average is conducted, the atomic stress $\sigma_{xx}$ computed by Eq. (3) with A-A dividing plane is zero, but $\sigma_{xx}$ will take a finite positive value if B-B dividing plane is used, so the total average $\sigma_{xx}$ of the system is positive. Although this is only a special moment of a special system, this incorrectness of Eq. (3) is actually universal since the tensile bonds in the atomic system are always longer than the compressed ones, and the arbitrary chosen dividing plane will more likely cut the tensile bonds than compressed bonds if their numbers are statistically identical for a stress free atomic system. This error in stress computation can be easily corrected by Eq. (2), which actually averages Eq. (3) over a period of time $\Delta t$ much longer than the shortest intrinsic vibration period. It should be pointed out that the Lagrangian dividing should not be changed in this average period $\Delta t$. Still using Fig. 5 as the example, the system is divided into two material sets by B-B cut at this moment and there are four bonds on the interface of these two sets. It is obvious that the average internal force of each bond over a sufficiently large $\Delta t$ is zero, therefore the $\sigma_{xx}$ from Eq. (2) becomes vanishing and correct. Similar to the spatial average, the temporal averaging period $\Delta t$ should also be microscopically large enough to obtain statistically meaningful quantities, while macroscopically short for capturing the time-dependent mechanical behaviors in continuum mechanics.

The necessity of temporal average can also be understood by noting the difference between continuum mechanics and discrete atomic mechanics. For an atomic system at a finite temperature and in thermodynamic equilibrium, all atoms keep in relative movements, which implies that the balance among internal forces and external forced usually can not be reached at an instantaneous moment, therefore a simple spatial average of internal forces can not yield a correct Cauchy stress as shown in Fig. 5. However, the balance among various forces can be statistically satisfied over a microscopically large period, which corresponds to the static and equilibrium state in continuum mechanics. Therefore, the configuration at a macroscopically instantaneous moment in continuum mechanics is essentially a temporal average of the discrete atomic system, and we have demonstrated above that the Lagrangian approach can produce the correct temporal average of internal forces, i.e. Cauchy stress.

Another issue should be mentioned is how to properly divide the volume element into two Lagrangian sets as shown in Fig. 4. Although this cut can be performed at any moment, our suggestion is that dividing the atoms with their average positions $\langle r_i \rangle$ is a more proper choice for obtaining statistical averaging quantities, such as the stress.

**4. Lagrangian virial stress**

Equation (2) is completely accurate and suitable for any circumstance, which is the reason that we call it "fundamental definition". In the following, some reasonable approximations or simplifications are made for computational convenience. In particular, Eq. (2) can be approximated as

$$\boldsymbol{n} \cdot \boldsymbol{\sigma} = \left\langle \frac{1}{S} \sum_{i \in Lag\ Set\ L} \boldsymbol{f}_{iR}^{Lag} \right\rangle \approx \frac{1}{S} \sum_{i \in Lag\ Set\ L} \left\langle \boldsymbol{f}_{iR}^{Lag} \right\rangle, \tag{4}$$

since there are a large number of atoms in the representative volume element as shown in Fig. 4, and the cross-sectional area $S$ has very small fluctuation during the macroscopically short averaging time duration $\Delta t$.

For the atomic system with pair interatomic potentials, Eq. (4) can be further written as

$$\boldsymbol{n} \cdot \boldsymbol{\sigma} = \frac{1}{S} \sum_{i \in Lag\ Set\ L} \left\langle \boldsymbol{f}_{iR}^{Lag} \right\rangle = \frac{1}{S} \sum_{i \in Lag\ Set\ L} \sum_{j \in Lag\ Set\ R} \left\langle \boldsymbol{f}_{ij} \right\rangle. \tag{5}$$

This equation can be used to compute the atomic stress of a cuboid volume element with the length $l$ and cross-sectional area $S$ in Fig. 6. As suggested above, the atom positions in the figure are average ones over $\Delta t$. $\langle \boldsymbol{f}_{ij} \rangle$ is the average internal force on atom $i$ exerted by atom $j$, and $\langle r_{ij} \rangle = \langle r_j - r_i \rangle = \langle r_j \rangle - \langle r_i \rangle$ is the average relative position vector from atom $i$ to $j$. According to Eq. (5), the pure contribution of $\langle f_{ij} \rangle$ to the atomic stress components $\boldsymbol{n} \cdot \boldsymbol{\sigma}$ of the volume element can be computed as $\dfrac{\boldsymbol{n} \cdot \langle r_{ij} \rangle}{l} \dfrac{\langle f_{ij} \rangle}{S}$, because the stress components $\boldsymbol{n} \cdot \boldsymbol{\sigma}$ in the gray region in Fig. 6 is $\dfrac{\langle f_{ij} \rangle}{S}$ and zero elsewhere, and $\dfrac{\boldsymbol{n} \cdot \langle r_{ij} \rangle}{l}$ represents the volume portion of the gray region. Therefore, by accounting for all interatomic forces, the definition on the atomic stress $\boldsymbol{\sigma}$ becomes

$$\boldsymbol{n} \cdot \boldsymbol{\sigma} = \frac{1}{2} \sum_{i,j \neq i} \frac{\boldsymbol{n} \cdot \langle r_{ij} \rangle}{l} \frac{\langle f_{ij} \rangle}{S} = \boldsymbol{n} \cdot \left( \frac{1}{2V} \sum_{i,j \neq i} \langle r_{ij} \rangle \otimes \langle f_{ij} \rangle \right), \quad (6)$$

where $V$ is the volume of the element, $\otimes$ denotes the tensor product of two vectors, and the prefactor 1/2 is introduced to avoid the double count of interatomic forces. We then obtain the stress for the volume element,

$$\boldsymbol{\sigma} = \frac{1}{2V} \sum_{i,j \neq i} \langle r_{ij} \rangle \otimes \langle f_{ij} \rangle, \quad (7)$$

and we name it the *Lagrangian virial stress*. There is also the corresponding Lagrangian virial stress for atom $i$,

$$\boldsymbol{\Pi}_i = \frac{1}{2\Omega_i} \sum_{j \neq i} \langle r_{ij} \rangle \otimes \langle f_{ij} \rangle, \quad (8)$$

here $\Omega_i$ is the volume around atom $i$ and should satisfy $\sum_i \Omega_i = V$. Obviously, the Lagrangian virial stress defined in Eqs. (7) and (8) can correctly predict the Cauchy stress of the example shown in Fig. 5. It should be pointed out that a mistake made in some earlier works (e.g., Ref.[12]) is essentially to replace $\langle r_{ij} \rangle \otimes \langle f_{ij} \rangle$ by $\langle r_{ij} \otimes f_{ij} \rangle$ in Eq. (8). The difference between the two definitions is the correlation between $r_{ij}$ and $f_{ij}$. Taking the harmonic spring for example (see Fig.5), $f_{ij}$ is linear with $r_{ij}$, so they have a non-zero correlation. This is similar to our previous argument that tensile bonds are always longer than compressed bonds. However, under some circumstances, the correlation between $r_{ij}$ and $f_{ij}$ can be exactly cancelled by the velocity term in the usual Virial stress expression, which will be discussed in the following sections.

## 5. Fundamental Eulerian atomic stress

There are two ways to describe the deformation or movement of solids: Lagrangian and Eulerian approaches. In the previous sections, we obtain the fundamental Lagrangian atomic stress which is completely consistent with the Cauchy stress in continuum mechanics. Therefore, we will derive the fundamental Eulerian atomic stress from its Lagrangian counterpart. Different from the Lagrangian dividing plane that may move with the material, the Eulerian dividing plane shown in Fig. 7 is fixed with respect to the reference frame. To link with the Lagrangian description, the colors white and black are still used in Fig. 7 to denote two Lagrangian sets.

If no atom crosses the Eulerian dividing plane, i.e. the Eulerian sets are identical with Lagrangian ones,

$$f_{iR}^{Lag} = f_{iR}^{Eul}, \tag{9}$$

and therefore the Eulerian atomic stress takes the same expression as the Lagrangian definition Eq. (2).

Otherwise, the momentum change of the left Eulerian set due to the atom crossing must be taken into account. We use a simple example to illustrate the contribution from some atom movements. Figure 7 shows three typical moments of the volume element: at the moment $t$, white atom $i$ just crosses the dividing plane towards the right part with the velocity $v_i^{out}$; at the moment $t + \Delta t$, it comes back and crosses the dividing plane with the velocity $v_i^{in}$; Fig. 7b shows the intermediate atomic configuration between $t$ and $t + \Delta t$, and atom $i$ is subject to the interatomic forces from the left Lagrangian set (colored white) $f_{iL}^{Lag}$, the forces from the right Lagrangian set (colored black) $f_{iR}^{Lag}$, and the external force $f_i^{ext}$. The $f_i^{ext}$ includes the real force imposed by the environment outside the system and the initial force due to the non-inertial reference frame. According to the impulse-momentum theorem, the momentum change of atom $i$ can be expresses as

$$m_i \left( v_i^{in} - v_i^{out} \right) = \int_t^{t+\Delta t} f_{iR}^{Lag} dt + \int_t^{t+\Delta t} f_{iL}^{Lag} dt + \int_t^{t+\Delta t} f_i^{ext} dt . \qquad (10a)$$

For simplicity, we first focus on the inertial reference frame case without external forces and discuss other circumstances later. Equation (10a) then becomes

$$m_i \left( v_i^{in} - v_i^{out} \right) = \int_t^{t+\Delta t} f_{iR}^{Lag} dt + \int_t^{t+\Delta t} f_{iL}^{Lag} dt . \qquad (10b)$$

Since the stress is essentially an average of $f_{iR}^{Lag}$ as mentioned above, we rewrite Eq. (10b) as

$$\int_t^{t+\Delta t} f_{iR}^{Lag} dt = \int_t^{t+\Delta t} f_{Li}^{Eul} dt + m_i \left( v_i^{in} - v_i^{out} \right), \qquad (11)$$

here $f_{Li}^{Eul} = -f_{iL}^{Lag}$ is the force acting on the left Eulerian set by atom $i$ during $(t, t+\Delta t)$, and the superscript "Eul" represents Eulerian set in this paper.

To integrate all contributions of the atoms in the volume element to the atomic stress, substituting Eqs. (9) and (11) into the Lagrangian atomic stress definition Eq. (2) yields

$$\boldsymbol{n}\cdot\boldsymbol{\sigma} = \frac{1}{S\Delta t}\left(\int_{t}^{t+\Delta t}\sum_{i\in Set_L^{Eul}} \boldsymbol{f}_{iR}^{Eul}dt + \sum_{j}m_j\boldsymbol{v}_j^{in} - \sum_{k}m_k\boldsymbol{v}_k^{out}\right), \qquad (12)$$

where $\sum_{i\in Set_L^{Eul}} \boldsymbol{f}_{iR}^{Eul}$ represents the sum of the internal forces spanning over the Eulerian dividing plane. Similarly, we name the stress $\boldsymbol{\sigma}$ in Eq. (12) the *fundamental Eulerian atomic stress*. It should be emphasized that this definition is valid under the following conditions, which have been used in previous derivation.

**Atom conservation condition**: The atoms crossing the Eulerian dividing plane will finally go back to their original side from time to time. Otherwise there is only $\boldsymbol{v}_i^{out}$ and no $\boldsymbol{v}_i^{in}$, and we can not use the momentum change to compute the internal force in Eq. (10b). This is necessary condition that was also mentioned by Marc and McMillian in their review article on the virial theorem[20].

**No external body force condition**: This is a strong condition that can ensure correctly using the momentum change to estimate the internal forces since the influence of non-zero external forces is difficult to take into account. In practice, if the reference frame is an inertial one and there is no other external body force, this condition is satisfied. The requirement of the inertial reference frame for the Eulerian type stress is also explicitly pointed out in Murdoch's derivation[21].

If the above conditions are satisfied, for two-body interatomic potentials, we may obtain the Eulerian virial stress.

## 6. Eulerian virial stress: the classical virial stress

Figure 8 shows an Eulerian volume element for computing the average atomic stress. In the following, the fundamental Eulerian definition in Eq. (12) is used, and the average of the first term on the right hand side is similar to previous section, so we focus our attention on the velocity-related term. For an infinitesimal time period $\delta t$, atom $i$ in Fig. 8 will cross those dividing planes over the gray region with the width of $\boldsymbol{n} \cdot \boldsymbol{v}_i \delta t$. The contribution from this atom movement to the atomic stress components $\boldsymbol{n} \cdot \boldsymbol{\sigma}$ of the gray region is $-\frac{1}{S \delta t} m_i \boldsymbol{v}_i$ based on Eq. (12), and the contribution to the total average atomic stress of the volume element is $-\frac{\boldsymbol{n} \cdot \boldsymbol{v}_i \delta t}{l} \frac{1}{S \delta t} m_i \boldsymbol{v}_i = -\frac{m_i}{V} \boldsymbol{n} \cdot \boldsymbol{v}_i \otimes \boldsymbol{v}_i$, here $\frac{\boldsymbol{n} \cdot \boldsymbol{v}_i \delta t}{l}$ is the volume portion of the gray region. Together with the average internal forces part, the total *Eulerian virial stress* is

$$\boldsymbol{\sigma} = \frac{1}{V} \left( -\sum_i m_i \boldsymbol{v}_i \otimes \boldsymbol{v}_i + \frac{1}{2} \sum_{i,j \neq i} \boldsymbol{r}_{ij} \otimes \boldsymbol{f}_{ij} \right). \qquad (13)$$

The corresponding Eulerian virial stress for atom $i$ is

$$\boldsymbol{\Pi}_i = \frac{1}{\Omega_i} \left( -m_i \boldsymbol{v}_i \otimes \boldsymbol{v}_i + \frac{1}{2} \sum_{j \neq i} \boldsymbol{r}_{ij} \otimes \boldsymbol{f}_{ij} \right), \qquad (14)$$

where $\Omega_i$ is the volume around atom $i$ and should satisfy $\sum_i \Omega_i = V$. Obviously, Eqs. (13) and (14) are the classical definitions of the virial stress.

## 7. Discussions on the objectivities of various definition of the atomic stress

In the previous sections, two groups of definitions have been given, Lagrangian and Eulerian ones. It should be pointed out that Gao and Weiner[22] have employed both approaches in studying rubber elasticity. However, the purpose of this paper is to discuss their objectivities or applicabilities. It is noted that the Eulerian atomic stress, Eqs. (12)-(14), include the velocities of atoms. The velocities are not objective quantities, which depend on the movement of the reference frame, and in different reference frames, the velocities will be different such that the Eulerian atomic stress will also be different. Therefore, the objective atomic stress in Eulerian approach can only be obtained by correctly choosing the reference frame. Two requirements for this selection have been given in the previous section, namely the *atom conservation condition* which requires that the atoms crossing the Eulerian dividing plane will finally go back to their original side from time to time, and the *no external body force condition*. In the following, we will use several examples to demonstrate the necessity of these two conditions.

Figure 9 shows a volume element in which each atom is at its equilibrium lattice position with the same velocity $v$ with respect to the inertial reference frame. To satisfy the *Atom Conservation Condition* of Eulerian definition, a new reference frame with the velocity $v$ relative to the old one can be chosen to compute the atomic stress more objectively, and the atom stress is obviously zero. This treatment has been noted by many scholars (e.g., Refs.[10,17]), and they suggest modifying the classical virial stress as follows,

$$\sigma = \frac{1}{V}\left(-\sum_i m_i (v_i - \bar{v}) \otimes (v_i - \bar{v}) + \frac{1}{2}\sum_{i,j \neq i} r_{ij} \otimes f_{ij}\right), \quad (15)$$

where $\bar{v}$ is the average velocity of the atoms in the local volume element, i.e. the velocity of the local reference frame. The introduction of $\bar{v}$ is essentially a partial Lagrangian treatment that may enhance the applicability of Eq. (15). However, as mentioned in the introduction part, $\bar{v}$ may also depend on the size of the volume element, and we use the following example to illustrate this point.

Figure 10 shows a volume element consisting of two groups of atoms (white and black) with two constant velocities, and no interatomic force is assumed. It has been known from the previous example that the atomic stress for the smaller volume element, no matter black or white atoms, is zero, so the total average atomic stress of the large volume element in Fig. 10 should be zero as well. However, if the large volume element is used in Eq. (15), since the local average velocity of the large volume $\bar{v}$ is zero, the total average atomic stress $\sigma_{xx}$ would be compression, different from the stress obtained from the small volume element. This mistake can also be attributed to the breaking of the *Atom Conservation Condition* of the reference frame for large volume element, i.e., once the atoms leave this Eulerian volume element, they will never return.

An example violating the *no external body force condition* is shown in Fig. 2, and has been discussed in the introduction. It is found that the nonzero external body forces induced by the alternative electric field $E$ may lead to the non-vanishing classical virial stress, while the correct Cauchy stress should be zero according to the balance of forces. It should be pointed out the *no external body force condition* is only a strong condition, and is easily used to check the applicability of the classical virial stress. In the previous sections, the classical virial stress is derived from Eq.(10b), in which the influence of the external forces $f_i^{ext}$ on the momentum change is not included. If $f_i^{ext}$ is taken into account in the derivation, Eqs. (13) and (14) will have an additional term (its value in the example of Fig.2 is obviously not zero), and the validness of the classical virial stress therefore requires the vanishing of this term. Of course, the *no external body force condition* can meet this requirement, but there may exist some other situations, in which the external body forces are nonzero but their effects can be cancelled out after the temporal and spatial averaging, i.e., this external force related term can also become zero. In this sense, the *no external body force condition* is just a strong condition. Moreover, it is found that the exact requirement for nonzero external body forces is not straightforwardly obtained and is not easily usable. We therefore suggest using *no external body force condition* as a strong requirement to ensure the correctness of the classical virial stress.

The Lagrangian atomic stresses in Eqs. (2), (7), (8), however, do not include any velocity in their expressions, and therefore do not have these velocity-related nonobjectivities. Moreover, in most cases, there is no requirement on external forces and on selecting the reference frame for the Lagrangian atomic stress such that both inertial and non-inertial reference frames are applicable, except one scenario as discussed below. When a material volume element undergoes rotation as shown in Fig. 11, the rotating local material reference frame should be used to correctly compute the Lagrangian atomic stress. However, because this reference frame is not inertial one, all Eulerian definitions of the stress would fail if the atoms have the relative velocities with respect to this rotating non-inertial reference frame.

The virial stress of single atom, Eq. (8) for Lagrangian definition and Eq. (14) for Eulerian definition, has a very simple expression and is easy to compute without choosing dividing plane, but may be wrong in some cases. Cheung and Yip[23] found that the virial stress can not correctly predict the surface stress. Here we use another example to show its limitation. For simplicity, we only investigate the static equilibrium case, i.e., no velocity is involved and Eqs. (8) and (14) become identical. Figure 12 shows a one-dimensional atomic chain with two types of interatomic interactions denoted by red and blue springs, respectively. The atomic chain is free and in equilibrium, but there is pre-tension $f^+$ in the red springs and pre-compression $f^- = -f^+$ in the blue springs. Atom 1 is at the right end, or the free surface, and its stress is supposed to be zero. However, from Eqs. (8) or (14), the stress of atom 1 is

$$\sigma_{xx}^{(1)} = \frac{1}{2(d/2)}(r_{12}f_{12} + r_{13}f_{13}) = \frac{1}{d}\left[(-d)(-f^-) + (-2d)(-f^+)\right] = f^+ > 0, \quad (16)$$

where $d$ is the atom spacing. In addition, the virial stress of atom 2 is

$$\sigma_{xx}^{(2)} = \frac{1}{2d}(r_{21}f_{21} + r_{23}f_{23}) = \frac{1}{2d}\left[df^- + (-d)(-f^-)\right] = f^- = -f^+ < 0. \quad (17)$$

Similarly, the virial stress of atom 3 is $f^+ > 0$. Furthermore, we use Eqs. (7) and (13), i.e. the virial stresses for volume elements, to compute the surface stress. The predicted surface stress is $f^+$ for the volume element including atom 1 only, $-f^+/3$ for one with atom 1 and 2, and $f^+/5$ for one with atom 1, 2 and 3, respectively. The computation above shows that the virial stress of atom 1 predicts a wrong non-vanishing surface stress, and this error may be reduced or corrected by averaging the stresses of more atoms. Therefore, the virial stress for a volume element, Eqs. (7) and (13), can yield more reasonable atomic stress than the virial stress of single atom Eqs. (8) and (14).

Actually, the fundamental Lagrangian atomic stress Eq. (2) can be used to compute the stress accurately under any circumstances since it does not have any approximation and simplification, which is just the reason why we name it with "fundamental". For the example shown in Fig. 12, if we divide the atomic chain into two material sets, the stress on the dividing plane would be $f^+ + f^- = 0$, accurately obtaining the vanishing surface stress.

Another advantage of the fundamental Lagrangian atomic stress Eq. (2) is its capability in computing the atomic stress for multi-body interatomic potentials. The virial stress is originally developed for two-body interatomic potentials. Recently, there are some efforts to extend the virial to multi-body potentials,[24,25] but the expressions are complex and the validness for general cases is not guaranteed as discussed above. We therefore suggest using Eq. (2) to compute the stress. One remained issue is how to calculate the internal force on atom $i$ exerted by the right Lagrangian set $\boldsymbol{f}_{iR}^{Lag}$ in Fig. 4 for multi-body potentials. Considering that the force on atom $i$ in any atomic system can be easily calculated, we compute $\boldsymbol{f}_{iR}^{Lag}$ from the difference of two forces

$$\boldsymbol{f}_{iR}^{Lag} = \boldsymbol{f}_i - \boldsymbol{f}_{iL}^{Lag}, \quad (18)$$

where $\boldsymbol{f}_i$ is the total force acting on atom $i$, and $\boldsymbol{f}_{iL}^{Lag}$ is the remaining total force acting on atom $i$ after imaginarily removing the right Lagrangian set.

Moreover, we noted that there sometimes exists nonuniqueness in previous works on the atomic stress for multi-body potentials (e.g., Refs[24,25]). In these works, the force on atom $i$ due to atom $j$ is determined as $f_{ij} = -\partial V / \partial r_{ij}$, where $V$ is the energy of an interatomic interaction. This expression may be vague in some cases. For example, one carbon atom is bonded to its four neighbor atoms in sp$^3$ atomic structures as shown in Fig. 13. The energy $V$ would depend on the distances among atoms and the angles between bonds based on Brenner's potentials[26,27]. It is easy to know that 9 independent geometry parameters can fully determine the relative positions of this structure, but there are 10 distances among 5 atoms ($r_{ij}$, $r_{ik}$, $r_{il}$, $r_{im}$, $r_{jk}$, $r_{jl}$, $r_{jm}$, $r_{kl}$, $r_{km}$, $r_{lm}$). Therefore, the internal force $f_{ij} = -\partial V / \partial r_{ij}$ is not unique and depends on which 9 distances are chosen. The virial stress Eq. (14) may also not be uniquely determined. However, the strategy suggested in our paper can successfully avoid these nonuniquenesses and has clear physical meaning.

## 8. Summary

The fundamental Lagrangian atomic stress in Eq. (2) is equal to the Cauchy stress under any circumstances since both definitions are essentially identical, and the temporal average in the definition must be performed over a period much longer than the shortest intrinsic vibration period, and the atomic stress suggested by Zhou[12] is incorrect due to the neglect of this temporal average. We proposed a Lagrangian virial stress Eqs (7) and (8) which possess simple virial forms and are easy to perform, while they do not involve the velocities. The corresponding Eulerian atom stresses are also derived, and the classical virial stress is actually the Eulerian virial stress. All these Eulerian stresses can be valid only when the impulse-momentum theorem is applicable to estimate the internal forces, such as the *atom conservation condition* and the *no external body force condition*. The virial stress of single atom may yield a wrong prediction of the stress, and the error can be reduced by averaging the atomic stress in a larger volume element. By contrast, there is no any error for the fundamental Lagrangian atomic stress. The atomic stress of the atomic system with multi-body interactions can be computed according to this fundamental Lagrangian definition as well.

**Acknowledgements**

BL acknowledges the support from National Natural Science Foundation of China (Grant Nos. 10542001, 10702034, 10732050) and National Basic Research Program of China (973 Program), 2007CB936803.

Figure captions

Figure 1. Snapshots of the dynamical configurations of an atomic bar with the stretching forces suddenly released.

Figure 2. Schematic of a sparse ion system under alternative electric field $E$.

Figure 3. Definition of Cauchy stress.

Figure 4. A Lagrangian volume element for computing the fundamental Lagrangian atomic stress. Open circles and solid dots represent the left and right Lagrangian sets, respectively.

Figure 5. A snapshot of an atomic system consisting of four one-dimensional atomic chains. A-A and B-B are two dividing planes for computing the atomic stress.

Figure 6. A volume element for computing the Lagrangian virial stress.

Figure 7. Three typical snapshots of an Eulerian volume element for computing the fundamental Eulerian atomic stress. A dividing plane is fixed with respect to the inertial reference frame.

Figure 8. A volume element for illustrating the Eulerian virial stress.

Figure 9. A volume element with the atoms moving at the same velocity. All atoms are at their equilibrium lattice positions and there is no interatomic force.

Figure 10. An atomic system with two groups of atoms (black and white atoms) moving at the same velocity but along opposite directions. Two volume elements denoted by the dashed line boxes are used to compute the atomic stress.

Figure 11. A local material volume element in rotation. The Lagrangian atomic stress

can be correctly obtained by choosing the reference frame with the same rotation velocity.

Figure 12. A one-dimensional atomic chain in static equilibrium with two types of interatomic potentials.

Figure 13. Schematic of local $sp^3$ atomic structure for carbon.

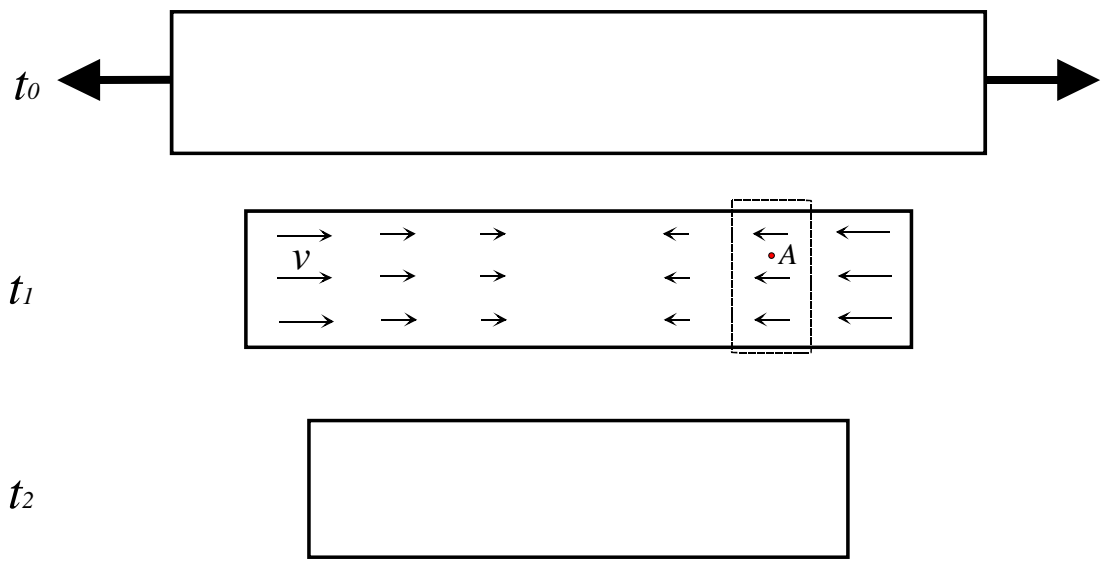

Figure 1. Snapshots of the dynamical configurations of an atomic bar with the stretching forces suddenly released.

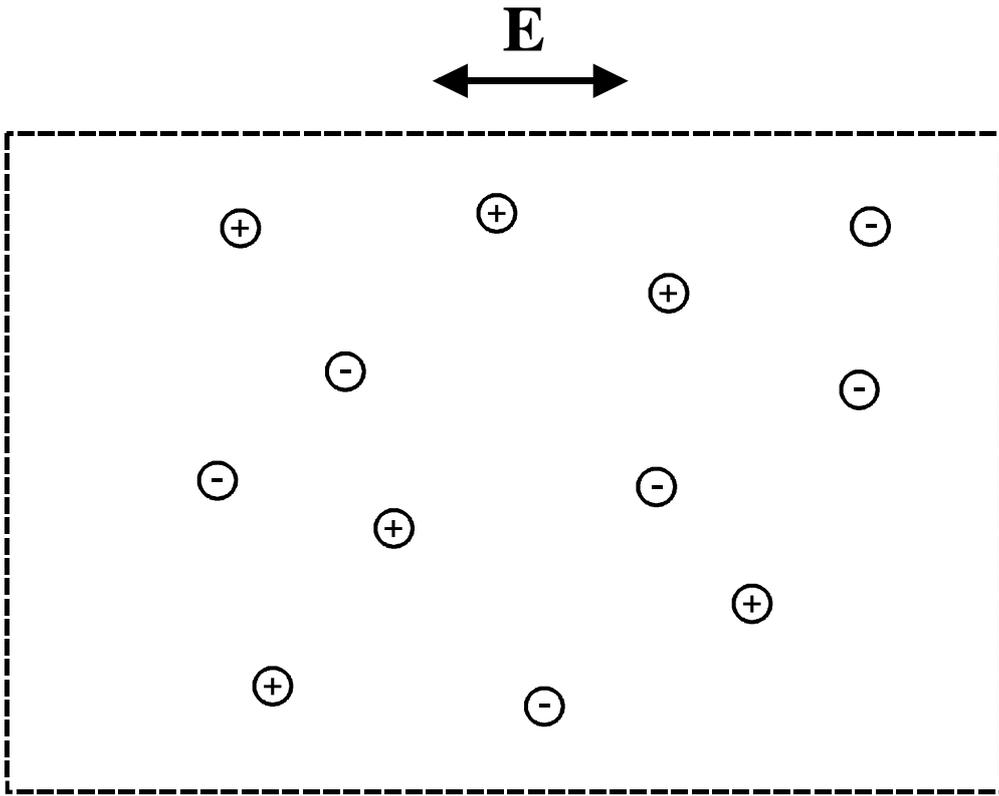

Figure 2. Schematic of a sparse ion system under alternative electric field $E$.

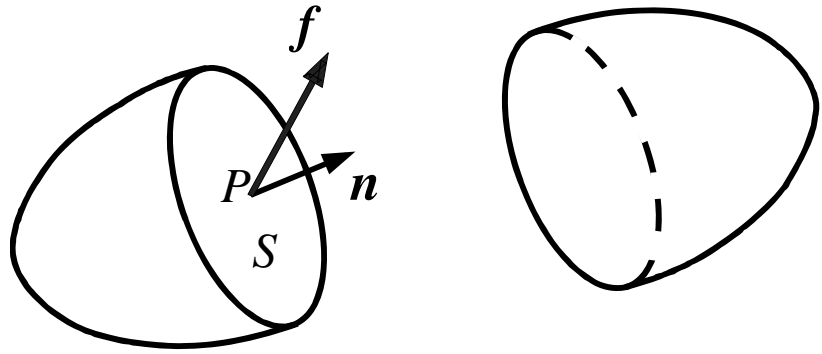

Figure 3. Definition of Cauchy stress.

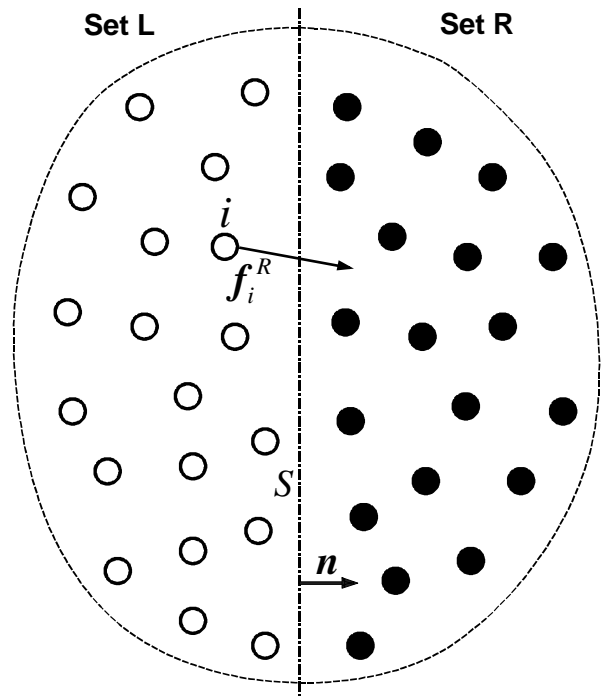

Figure 4. A Lagrangian volume element for computing the fundamental Lagrangian atomic stress. Open circles and solid dots represent the left and right Lagrangian sets, respectively.

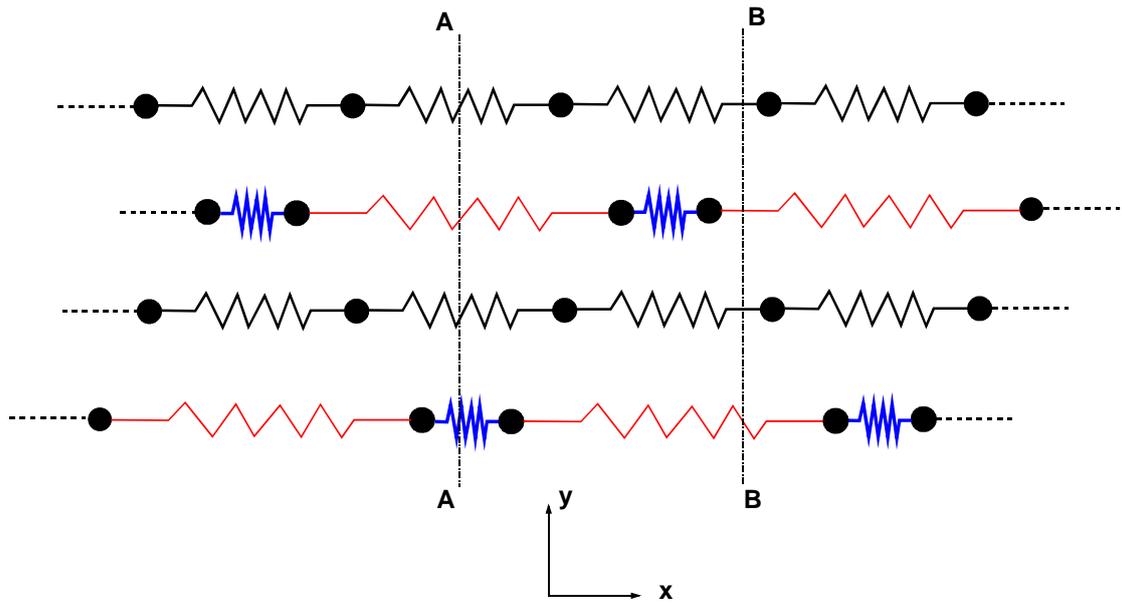

Figure 5. A snapshot of an atomic system consisting of four one-dimensional atomic chains. A-A and B-B are two dividing planes for computing the atomic stress.

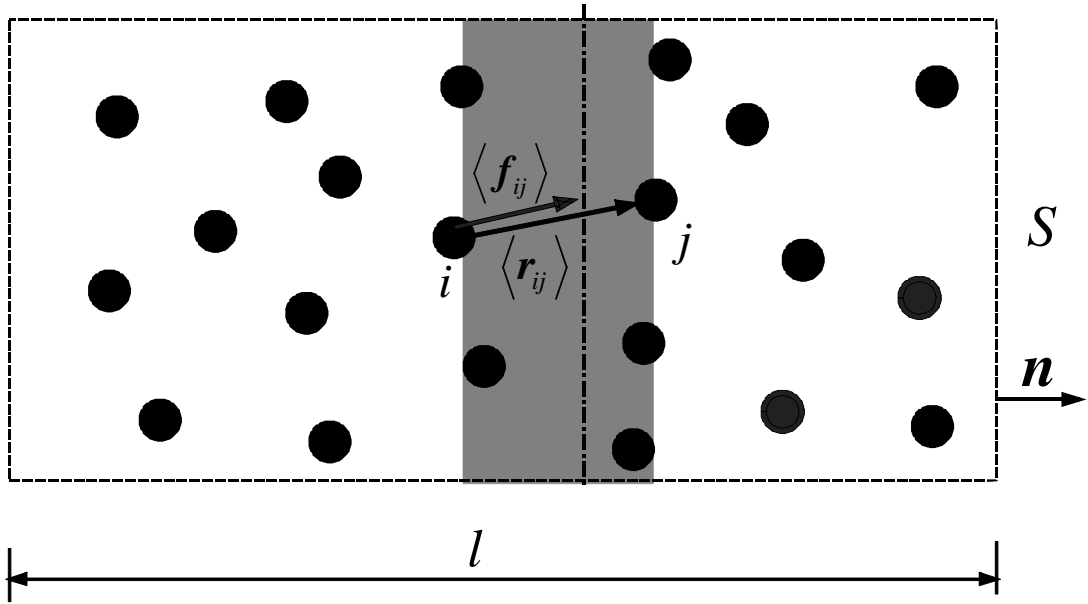

Figure 6. A volume element for computing the Lagrangian virial stress.

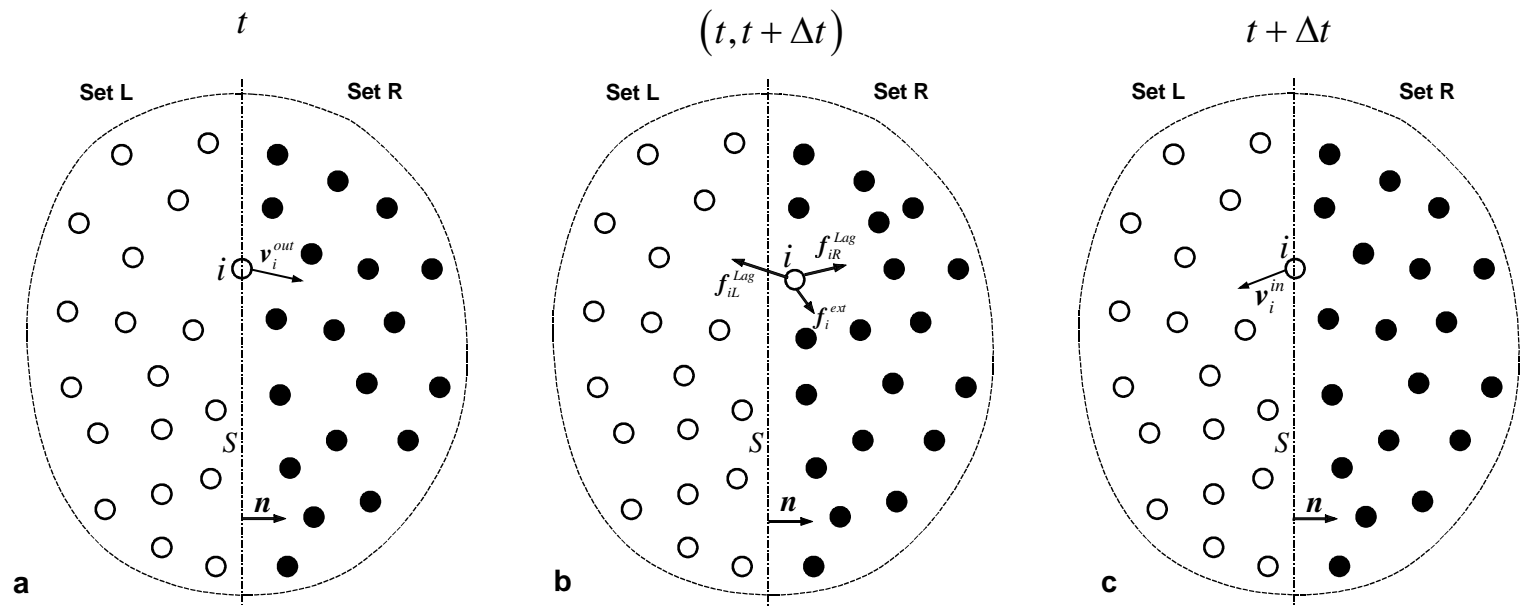

Figure 7. Three typical snapshots of an Eulerian volume element for computing the fundamental Eulerian atomic stress. A dividing plane is fixed with respect to the inertial reference frame.

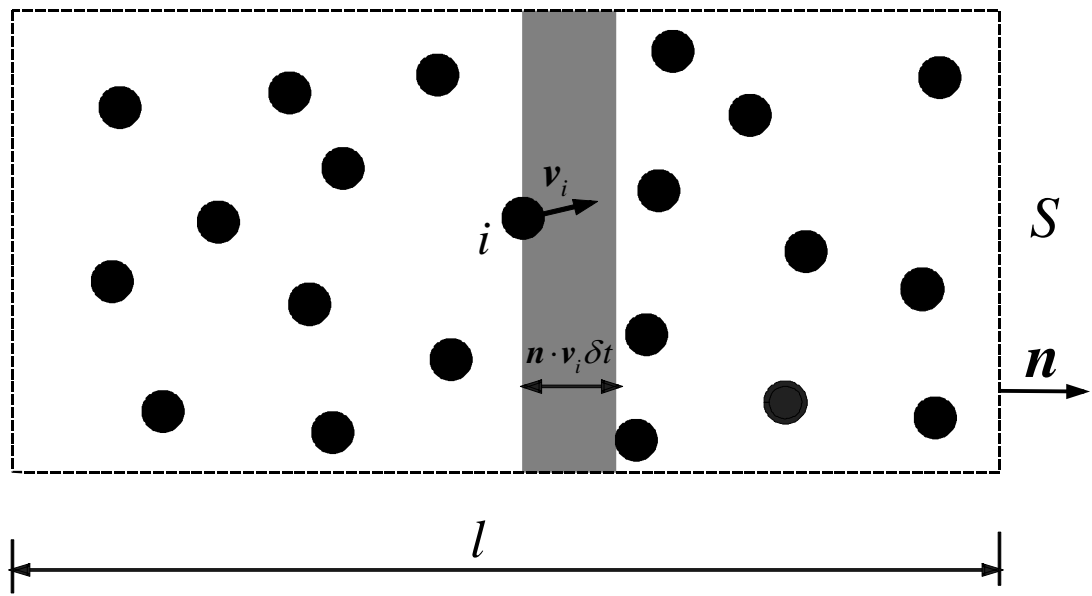

Figure 8. A volume element for illustrating the Eulerian virial stress.

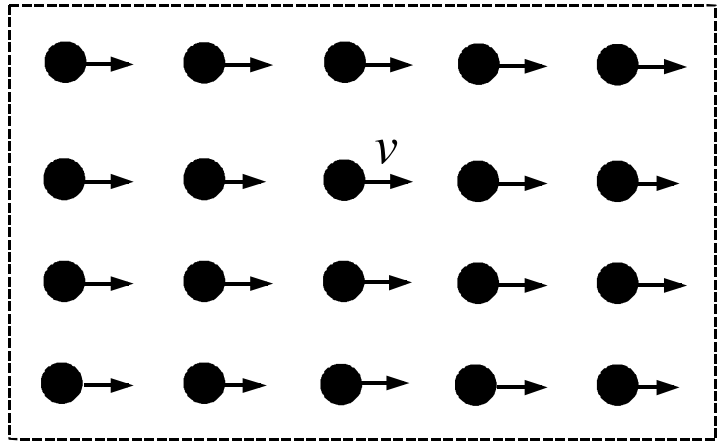

Figure 9. A volume element with the atoms moving at the same velocity. All atoms are at their equilibrium lattice positions and there is no interatomic force.

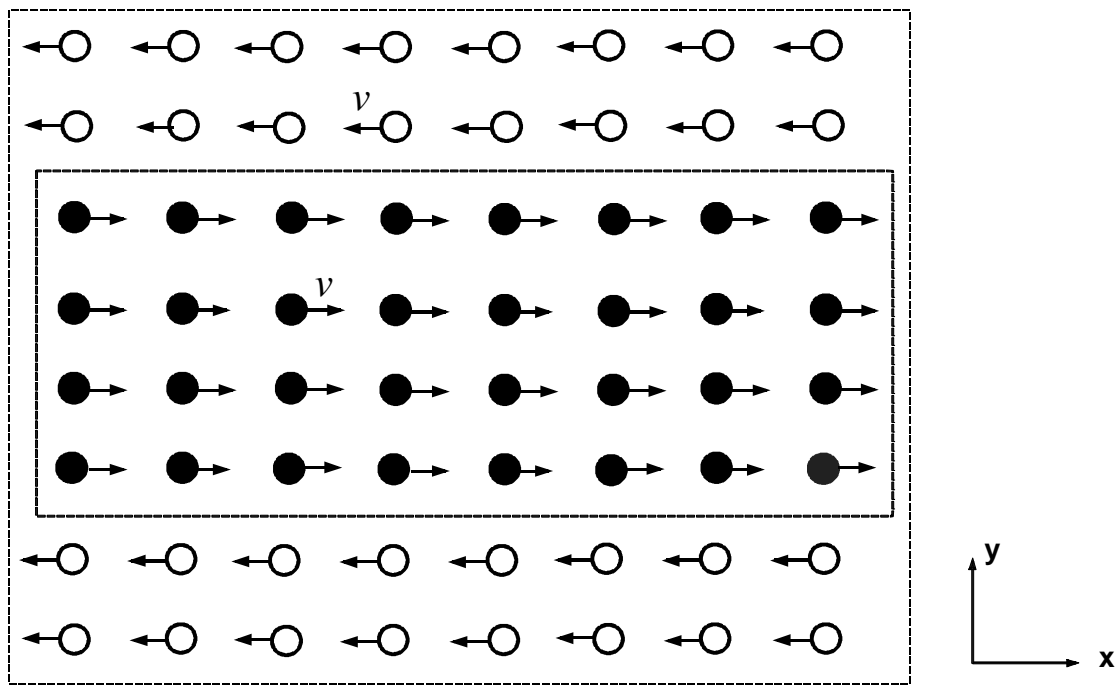

Figure 10. An atomic system with two groups of atoms (black and white atoms) moving at the same velocity but along opposite directions. Two volume elements denoted by the dashed line boxes are used to compute the atomic stress.

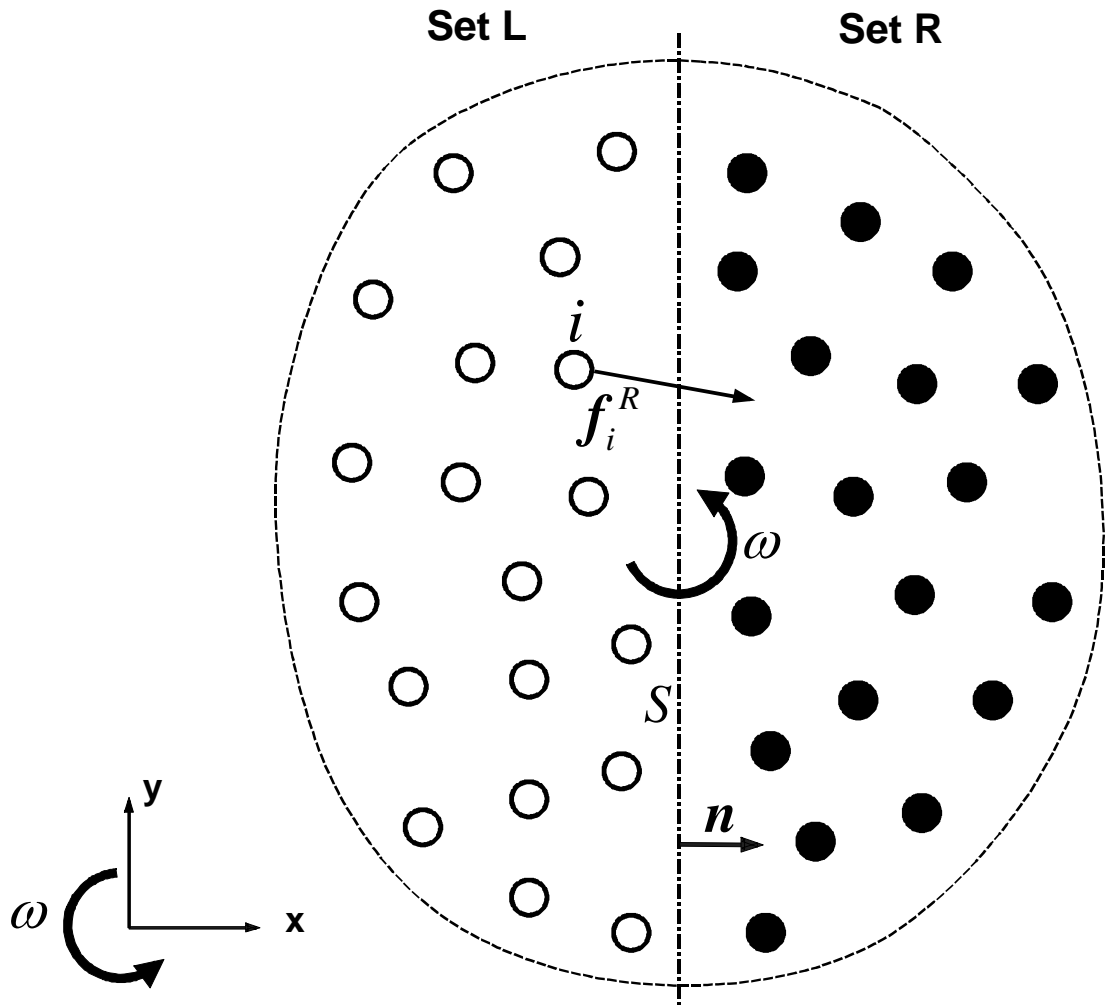

Figure 11. A local material volume element in rotation. The Lagrangian atomic stress can be correctly obtained by choosing the reference frame with the same rotation velocity.

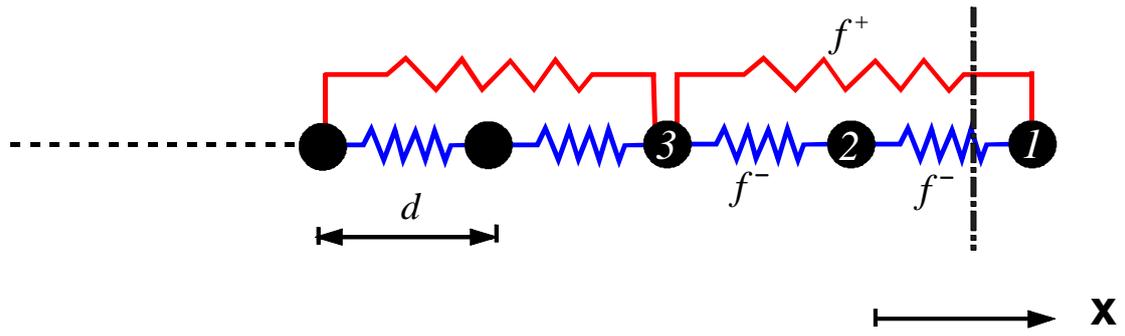

Figure 12. A one-dimensional atomic chain in static equilibrium with two types of interatomic potentials.

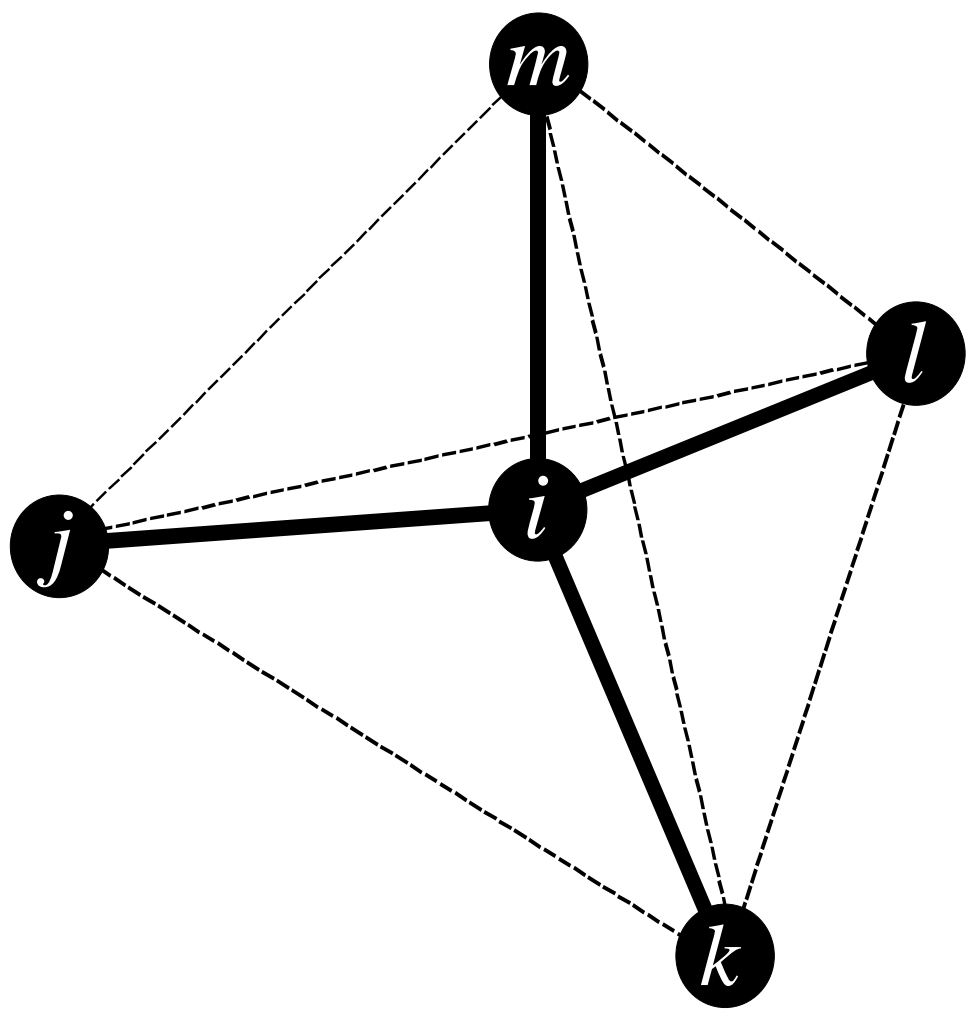

Figure 13. Schematic of local sp$^3$ atomic structure for carbon.